\documentclass[12pt]{iopart}
\pdfoutput=1
\usepackage[pdftex]{graphicx}

\begin{document}

\title[Gradient echo memory in an ultra-high optical depth cold atomic ensemble]{Gradient echo memory in an ultra-high optical depth cold atomic ensemble}

\author{B. M. Sparkes$^1$, J. Bernu$^1$,  M. Hosseini$^1$, J. Geng$^{1,2}$, Q. Glorieux$^{1,3}$, P. A. Altin$^4$, P. K. Lam$^1$, N. P. Robins$^4$, and B. C. Buchler$^1$.}

\address{$^1$ Centre for Quantum Computation and Communication Technology, Department of Quantum Science, Research School of Physics and Engineering, The Australian National University, Canberra, ACT 0200, Australia}
\address{$^2$ Quantum Institute for Light and Atoms, Department of Physics, East China Normal University, Shanghai 200062, P. R. China}
\address{$^3$ Joint Quantum Institute, University of Maryland, College Park, Maryland 20742, USA}
\address{$^4$ Quantum Sensors Lab, Department of Quantum Science, Australian National University, Canberra, ACT 0200, Australia}
\ead{ping.lam@anu.edu.au}

\begin{abstract}
Quantum memories are an integral component of quantum repeaters - devices that will allow the extension of quantum key distribution to communication ranges beyond that permissible by passive transmission. A quantum memory for this application needs to be highly efficient and have coherence times approaching a millisecond. Here we report on work towards this goal, with the development of a $^{87}$Rb magneto-optical trap with a peak optical depth of 1000 for the D2 $F=2 \rightarrow F'=3$ transition using spatial and temporal dark spots. With this purpose-built cold atomic ensemble we implemented the gradient echo memory (GEM) scheme on the D1 line. Our data shows a memory efficiency of $80\pm 2$\% and coherence times up to 195~$\mu$s, which is a factor of four greater than previous GEM experiments implemented in warm vapour cells.
\end{abstract}

\section{Introduction}
\label{sec:cagem_introduction}

The development of quantum key distribution (QKD) over the last few decades \cite{Bennett1984} brings with it the promise of secure global communication. Although there have been some heroic efforts, the current maximum distance for QKD is 260 km \cite{Wang2012a}, achieved using optical fibres. Beyond this length scale, the loss of photons from the quantum channel makes provably secure single-step QKD impossible. One way of overcoming the distance limitation is to use a quantum repeater (QR), a device based on entanglement swapping along nodes placed between the start and end locations of the signal \cite{Duan2001}. A working QR will require two key components: a source of entanglement; and a quantum memory to store the entanglement at the nodes \cite{Hammerer2010a,Sangouard2011a}. Storage is required because the generation, detection and distillation of entanglement is not deterministic. In order to establish entanglement between repeater nodes, entanglement has to be stored and recalled on demand.

For use in a QR, a quantum memory must meet a number of requirements.  The efficiency of the memory must be as high as possible, ideally approaching unity.  The efficiency is important for bit rates: inefficient storage means that the chance of generating entanglement in any given attempt is lower and it will take longer to complete a swapping operation.  The storage time must be long as this will limit the maximum distance between nodes in a QR network. For instance, if the distance between nodes were to be 100~km, then more than half a millisecond of storage is required. A high storage bandwidth is also desirable. The bandwidth will limit the kind of entanglement that can be stored in the memory. The most common sources of entangled photons are based on spontaneous parametric downconversion and these sources have a wide bandwidth. A narrow bandwidth memory will therefore require new sources of entanglement \cite{Wolfgramm2008,Zhang2011b,Glorieux2012a}. A multimode quantum memory, in which multiple pieces of entanglement can be stored simultaneously,  would allow for faster bit-rates in QRs, and there are QR protocols that can use multi-mode memories to improve QR designs \cite{Simon2007,Collins2007}. The product of the storage time and the bandwidth, known as the delay-bandwidth product (DBP) gives an indication of the number of pieces of information that can usefully be stored in a multimode memory.

The challenge for experimentalists developing quantum memory prototypes is to realise all these properties in a single system. The practicalities of achieving this are intertwined with the choice of atomic ensemble.  The current menu has a choice of three: warm gaseous atomic ensembles \cite{Hosseini2011c,Reim2010a,Reim2011}; solid state systems \cite{Longdell2005,Hedges2010,Amari2010,Bonarota2011}; and cold gaseous atomic ensembles, which will be the focus of this work.

No matter what the storage medium, the key to achieving high efficiencies is to have a high optical depth (OD). This is normally achieved by having as high a density of absorbers in the interaction volume as possible. To date the highest efficiency of any unconditioned quantum memory prototype is 87\% \cite{Hosseini2011c}. This was achieved using the gradient echo memory (GEM) scheme in a warm atomic vapour. This system used a 20-cm-long cell of isotopically pure $^{87}$Rb and 0.07~kPa of Kr buffer gas and had a resonant OD in the thousands \cite{Hosseini2012}. The thermal motion of the atoms is the primary obstacle to increasing the storage lifetime in warm vapour \cite{Hosseini2012, Higginbottom2012a}, which was found to decay with a time constant of around 50~$\mu$s. 

Reducing atomic motion is not the only way to increase storage times - it depends on the storage protocol. For example, it has been demonstrated that long storage times can be obtained in an electromagnetically-induced transparency (EIT)-based warm memory by using high buffer gas pressure \cite{Phillips2008}.   In a GEM scheme, however, high buffer gas pressure increases collisional broadening and absorption of the coupling beam \cite{Hosseini2012}. The effect of transverse diffusion can be minimised by increasing the interaction volume and using anti-relaxation coatings to minimise inelastic wall collisions \cite{Balabas2010,Balabas2010a}. Although this technique would reduce transverse diffusion, the longitudinal diffusion would eventually limit coherence times. This is because, owing to the longitudinal frequency gradient needed for GEM, longitudinal diffusion introduces random frequency changes to the atoms during storage, which lowers the memory fidelity.

One obvious way to overcome the impact of atomic motion is to use slow-moving atoms. Laser cooling provides an efficient way to achieve this, and cold atom memories have been the subject of many experiments aiming for longer memory storage times. The majority of these cold atom experiments have used EIT as the storage method. A coherence time of 540~ms was achieved by cooling approximately $3 \cdot 10^6$ Na atoms into a Bose-Einstein condensate (BEC) and placing them in a dipole trap \cite{Zhang2009}. The maximum memory efficiency in this ensemble was, however, just a few percent.  A much longer coherence time of 3~s was achieved using a cross-beam dipole trap of $3\cdot10^{5}$ $^{87}$Rb atoms \cite{Sagi2010}, although no memory was demonstrated in this experiment.  

By optimising for higher atom number and therefore higher OD, high efficiency is also possible in cold atom systems, although this can come at the expense of long storage time due to higher temperatures. A memory efficiency of 50\% was achieved using a 2D magneto-optical trap (MOT) of $^{85}$Rb with an OD $<140$, and a coherence time of a few hundred nanoseconds \cite{Zhang2011}. Similar efficiencies were achieved for single-photon storage in a $^{87}$Rb BEC in a cross-beam dipole trap containing $1.2 \cdot 10^6$ atoms \cite{Riedl2012}. The coherence time for this experiment was approximately 500~$\mu$s. More recently, 78\% efficiency and a coherence time of 98~$\mu$s \cite{Chen2012} has been achieved using a MOT with $10^9$ atoms and an OD of 160. It has also been demonstrated that, by creating a MOT inside a ring cavity, single photons can be generated efficiently with the atomic coherence being maintained for up to 3~ms \cite{Bao2012}. Finally, experiments have also been done showing the interaction of quantum states of light with cold-atom-based EIT. Continuous-wave sideband squeezing \cite{Arikawa2010} and discrete variable entanglement \cite{Zhang2011b} have both been investigated.

In this paper we introduce a cold atom realisation of the GEM scheme.  This protocol has resulted in the highest efficiencies in both warm vapour \cite{Hosseini2011c} and solid-state memory systems \cite{Hedges2010}. In both these realisations, it was shown that GEM adds no noise to the output. It has also been shown to be temporally \cite{Hosseini2011c}, spatially \cite{Glorieux2012,Higginbottom2012a}, and spectrally multi-mode \cite{Sparkes2012}. Given that the warm vapour GEM storage time is limited by atomic motion, a cold atom version of this protocol is well motivated. To achieve this we first had to develop an ultra-high optical depth atomic source to allow for high efficiency recall.

The body of the paper is split into two sections: in Section~\ref{sec:cagem_mot} we will present the experimental details of our cold atomic source; and in Section~\ref{sec:cagem_gem} we will present the basic theory behind GEM and a high efficiency demonstration of GEM using cold atoms. In Section~\ref{sec:cagem_discussion} we discuss possible improvements and other experiments that could benefit from a high optical depth atomic source. Conclusions are presented in Section~\ref{sec:cagem_conclusions}.

\section{The Cold Atomic Source}
\label{sec:cagem_mot}

\begin{figure}[!ht]
\begin{center}
\includegraphics[width=\columnwidth]
{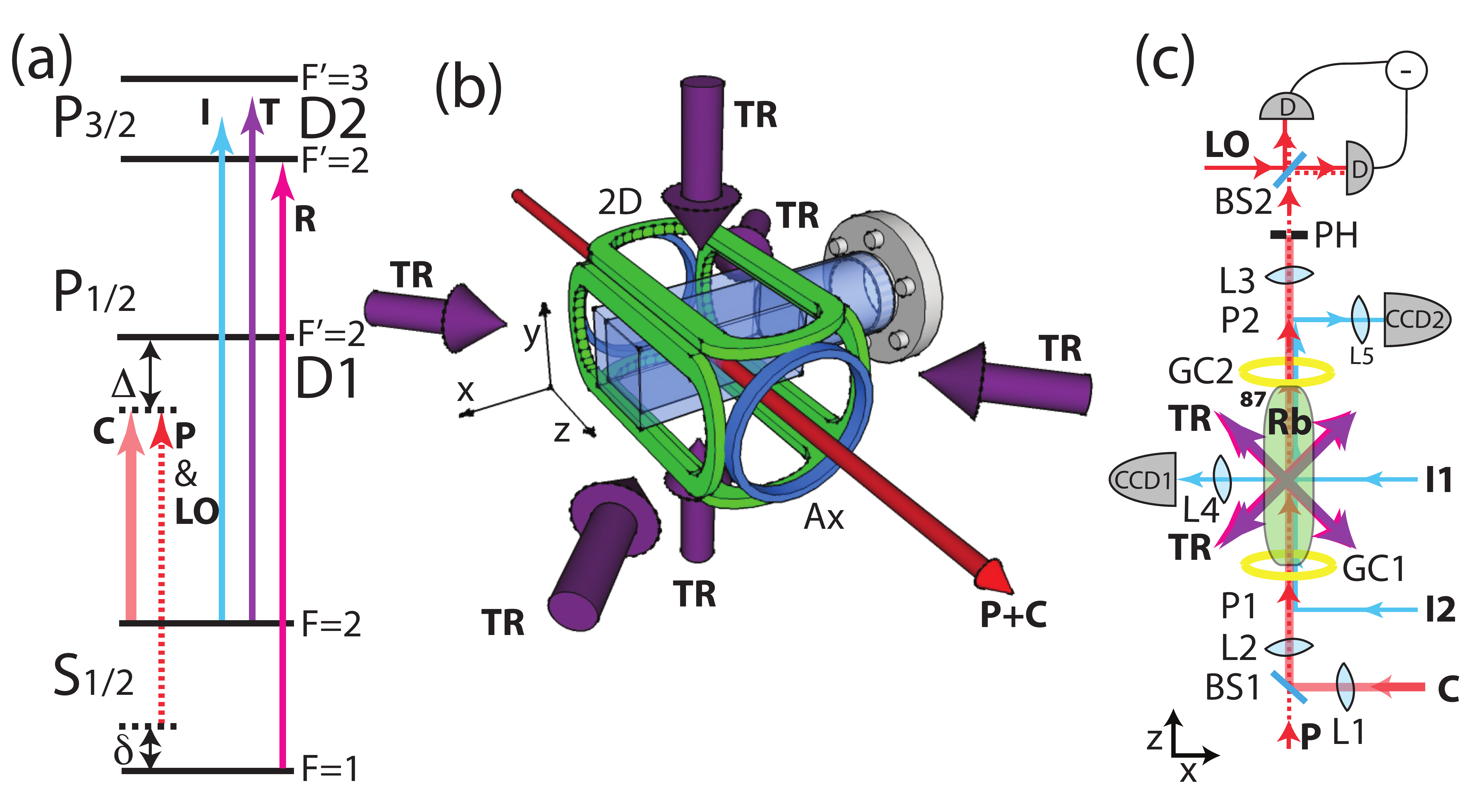}
\caption{Experimental set-up. (a) Energy level diagram for all fields present: C - coupling; P - probe;  LO - local oscillator; I - Imaging; T - trapping; R - repump. Also shown are one- and two-photon detunings for probe and coupling fields ($\Delta$ and $\delta$ respectively). (b) 3D representation of MOT configuration: TR - trapping and repump fields; 2D - $x$ and $y$ MOT coils (green); Ax - axial ($z$) coils (blue). (c) 2D schematic diagram of experiment: $^{87}$Rb - atomic ensemble; BS - 50:50 beam-splitter; L - lens; D - photo-diode detector; I1(2) - imaging beam 1(2); CCD - charged-coupled device camera; PH - pin-hole; GC - GEM coils; P1 and P2 - positions for inserting mirrors for axial ($z$) imaging with I2. MOT coil configuration not shown, neither is vertical ($y$) MOT beam.}
\label{fig:cagem_setup}
\end{center}
\end{figure}

The MOT is a workhorse of modern atomic physics as it offers the robust and efficient collection, cooling and storage of cold atoms.  A typical MOT operating on $^{87}$Rb will quickly cool (tens of milliseconds) and trap $10^8$ atoms from a dilute thermal background gas at densities of approximately $10^{10} $~atoms/cm$^3$ and temperatures on the order of 100~$\mu$K.

For quantum memory applications we aim to reach low temperatures and the very large ODs required for high efficiencies.  Because optical depth is related to the integrated absorption of photons through a sample, there are a a number of clear ways to increase the OD: increase the atom number; increase the density; or increase the length of the atom cloud. Low temperature is also important as the thermal diffusion of atoms is a limiting factor for the memory lifetime. In addition, the memory cannot be operated while either the magnetic or optical trapping fields are active. Cold atoms expand ballistically when released from a MOT, with a typically Gaussian velocity distribution related to their kinetic temperature. The density of a sample decreases as it expands, reducing the optical depth. A lower temperature means a slower expansion of the cloud and therefore higher densities after switch off of the trap.  

In the following we present a MOT that achieves a peak OD of over 1000 at a temperature of 200~$\mu$K.  We obtain this result by optimising both the static loading of the MOT through geometry, a spatial dark spot and optical de-pumping, followed by a compression phase using a temporal dark spot \cite{DePue2000}.  After describing the application of these techniques, we present a characterisation of the system using near resonant absorption imaging.

\subsection{Loading Phase}
\label{sec:cagem_static}
Our $^{87}$Rb MOT is in a three beam retro-reflection configuration.  The trapping and cooling laser has a total power of 400~mW after spatial filtering and is red-detuned by 30~MHz from the D2 $F=2 \rightarrow F'=3$ transition for the loading of the MOT. The repumping field is on resonance with the D2 $F=1 \rightarrow F'=2$ transition, as shown in Figure~\ref{fig:cagem_setup}(a). Figure~\ref{fig:cagem_setup}(b) shows the MOT beam configuration, with trapping and repump beams being combined on a polarising beam-splitter (PBS) and then split further using another PBS (neither pictured) to create three cooling beams of approximately 3.5~cm diameter, all of which are then appropriately polarised and retro-reflected.

The optimal shape for our atomic ensemble is a cylinder along the direction of the memory beams to allow for maximum absorption of the probe. This can be achieved by using rectangular, rather than circular, quadrupole coils \cite{Lin2008,Peters2009} or using a 2D MOT configuration \cite{Zhang2011, Zhang2012}. To create this shape while still allowing easy access for the memory beams we use four elongated coils in a quadrupole configuration to create a 2D MOT in the $z$ direction (memory axis) and position the horizontal MOT beams at 45$^\circ$ to the long axis of the MOT. An extra set of axial coils in the $z$ direction creates 3D confinement.  The shape of the MOT can then be determined by the currents in the 2D and axial coils, as well as the relative intensities of the trapping fields.

For the loading phase, the cylindrically symmetric magnetic field gradient produced by the 2D MOT coils is 16~G/cm, and for the axial coils is 2~G/cm.  The $^{87}$Rb atoms are produced from a natural-mixture Rb dispenser inside a $100\times50\times50$ mm$^3$ single-sided, anti-reflection coated cell shown in Figure~\ref{fig:cagem_setup}(b). This cell is attached to a vacuum system consisting of a 70~L/s ion pump supplemented by a passive titanium sublimation pump; with the dispenser running in the cell we measure a background pressure at the ion pump of $1.5\cdot10^{-9}$~kPa.

The density in the trapped atomic state ($F=2$ here) is limited by reabsorption of fluorescence photons within the MOT (leading to an effective outwards radiation pressure \cite{Metcalf2003}).  By placing a dark spot of approximately 7.5~mm in diameter in the repump, atoms at the centre of the MOT are quickly pumped into the lower ground state ($F=1$) and become immune to this unwanted effect, allowing for a higher density of atoms in the centre of the trap, as first demonstrated in \cite{Ketterle1993}. 

We can collect over $10^{10}$ atoms in this configuration.  However, we find that the static MOT parameters that optimise atom number do not optimise density in the compression and cooling phase of the MOT, and we typically work with a sample of $4\cdot10^{9}$ atoms. 

\subsection{Compression Phase}
\label{sec:cagem_dynamic}

In this second stage of the ensemble preparation we utilise a temporal dark spot to transiently increase the density of the sample by simultaneously increasing the trapping laser and repump detunings and increasing the magnetic field gradient \cite{Lee1996,DePue2000,Petrich1994}.  Detuning the trapping laser creates some of the conditions for polarisation-gradient cooling (PGC) \cite{Dalibard1989}, which can be used to achieve much colder and denser ensembles than in a standard MOT.  We smoothly ramp the frequency of the trapping beam from 30 to 70~MHz below resonance, and the repump beam to 8 MHz~below resonance, over a period of 20~ms.  The 2D magnetic field gradient in the $x$ and $y$ directions is ramped up to 40~G/cm as the trapping and repump lasers are detuned. We do not ramp the axial field.

Finally, we optically pump the atoms into the desired ground state ($F=1$, see Figure~\ref{fig:cagem_setup}(a)) by simply turning off the repump 50~$\mu$s before the trapping beam.

\subsection{MOT Characterisation}
\label{sec:cagem_characterisation}
\begin{figure}
\begin{center}
\includegraphics[width=\columnwidth]{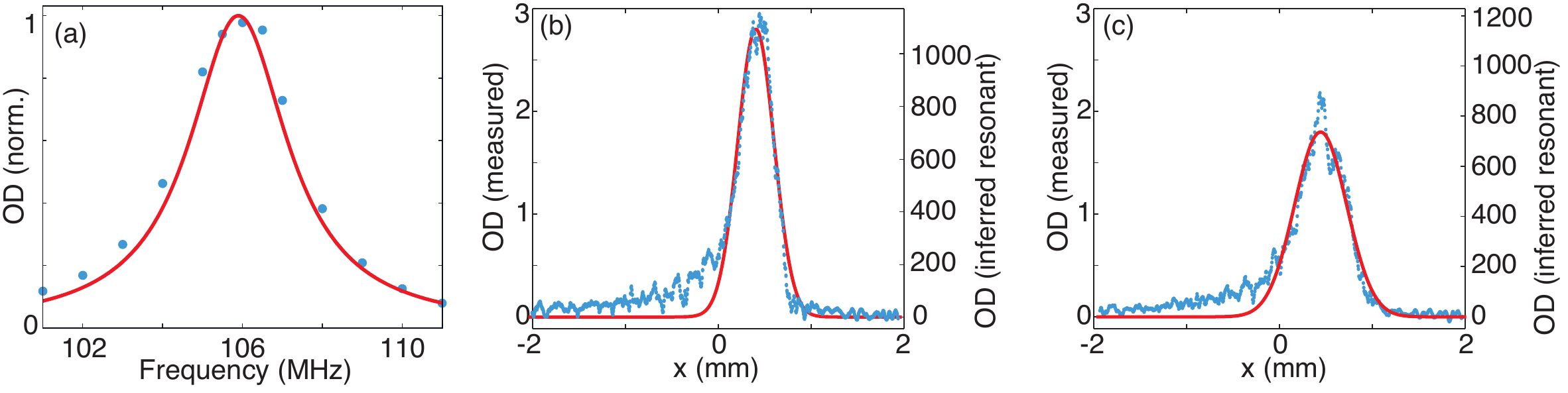}
\caption{OD characterisation. (a) Normalised OD as a function of imaging beam AOM frequency, with low atom number. (b) Averaged scaled (left-hand scale) and unscaled (right-hand scale) OD cross-sections, taken across 10 $y$-plane slices for seven traces, as a function of $x$ position for an imaging beam detuning of -60.8~MHz from resonance (OD scale factor $=410$) and (c) for +59.2~MHz (OD scale factor $=390$). Points represent data, red lines represent fits (Lorentzian in (a), Gaussian in (b) and (c)).}
\label{fig:cagem_odcal}
\end{center}
\end{figure}

Absorption imaging was used to optimise and characterise the MOT. The set-up and frequency of the imaging beams are shown in Figure~\ref{fig:cagem_setup}. We perform imaging both across (I1) and along (I2) the $z$ axis (in which case two mirrors are temporarily placed at locations P1 and P2).

Absorption imaging allows us to measure the OD of the MOT using Beer's law
\begin{equation}
\mathrm{OD} \simeq \mathrm{ln}(I_{t}/I_o),
\label{eq:cagem_od}
\end{equation}
where $I_t$ is the transmitted imaging beam intensity after passing through the MOT, and $I_o$ is the intensity measured without the MOT present. This is calculated for every point in the imaging plane.

As the absorption of light by atoms away from resonance will follow a Lorentzian decay, to be able to have a precise value for OD it is important to have a well calibrated line centre. This is especially important as one goes further off resonance as the relation between on-resonance OD and off-resonance OD depends on the one-photon detuning ($\Delta$) as follows:
\begin{equation}
\mathrm{OD}_{res} = \frac{\Delta^2+\gamma^2/4}{\gamma^2/4} \mathrm{OD}_{\Delta},
\label{eq:cagem_odscaling}
\end{equation}
where $\gamma$ is the excited state decay rate. For Rb $\gamma$ is approximately 6 MHz.

To measure $\Delta$ accurately we lowered the atom number in our trap until the OD did not saturate on resonance and plotted out the resonance curve as a function of detuning to accurately locate the line centre. This is shown in Figure~\ref{fig:cagem_odcal}(a).

\begin{figure}[!ht]
\begin{center}
\includegraphics[width=\columnwidth]
{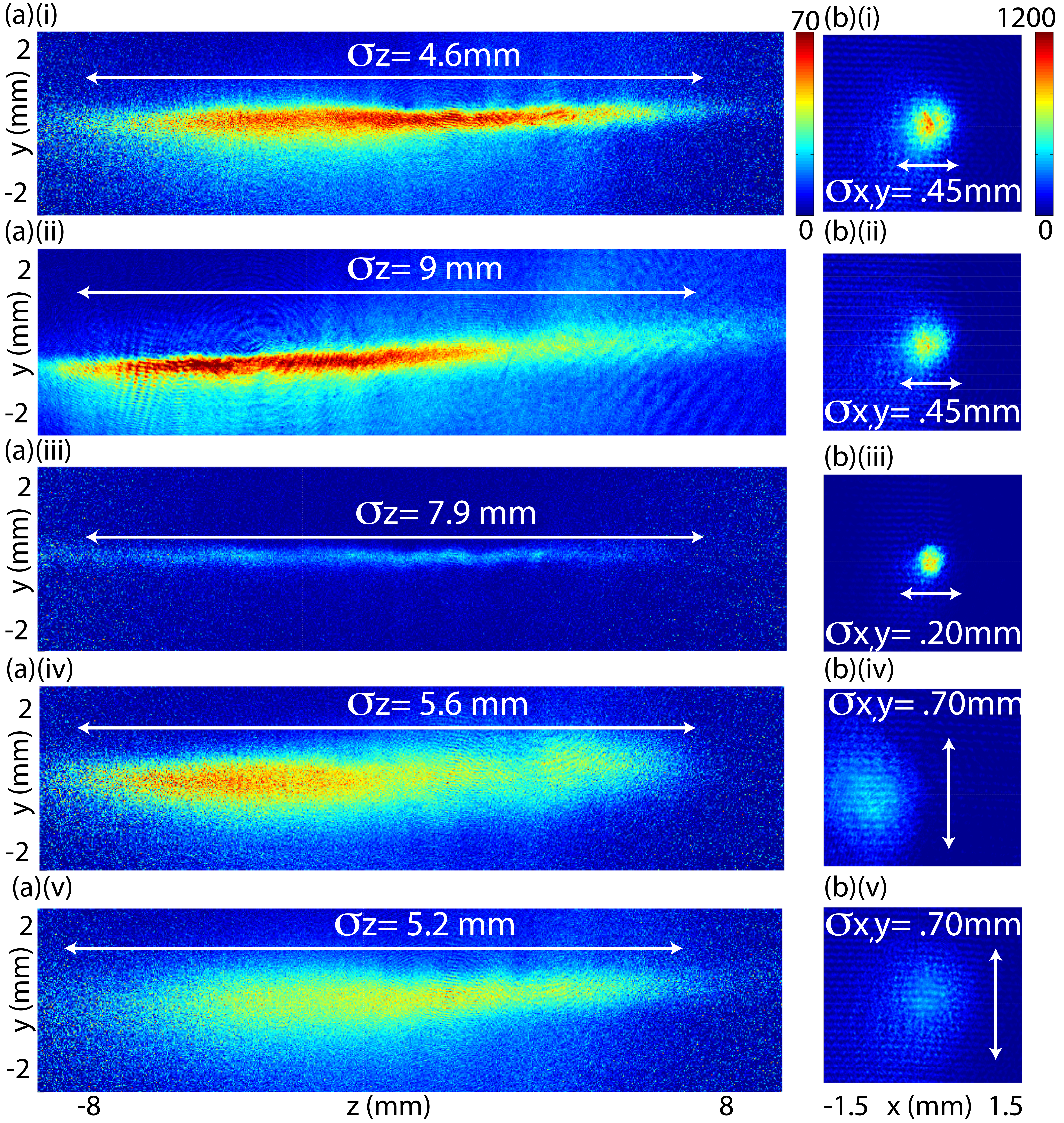}
\caption{MOT Characterisation Images. (a) Side-on and (b) front-on images of the atomic ensemble for various MOT parameters taken at 500 $\mu$s after fields turned off: (i) all optimisation protocols described in text; (ii) no dark spot; (iii) no axial gradient; (iv) no magnetic field compression; and (v) same as (i) but 3 ms after fields turned off. Colour bars on (a) traces show OD scales, $\sigma$ values show the standard deviation of Gaussian fits to ensemble.}
\label{fig:cagem_MOTzoo}
\end{center}
\end{figure}

Images of the MOT for various configurations are shown in Figure~\ref{fig:cagem_MOTzoo}. For these imaging runs, we used a 4.98~s load time followed by 20~ms of ramping fields as described in Section~\ref{sec:cagem_dynamic}. All fields were then turned off and an image of the MOT was taken 500~$\mu$s later (unless otherwise stated), with a comparison image being taken 150~ms later to obtain as precise a measure of $I_o$ as possible while ensuring no atoms were still present. As the imaging beam is on the closed D2 $F=2 \rightarrow F'=3$ transition, it was necessary to pump atoms back into the $F=2$ state before these images were taken. This was achieved with a 200~$\mu$s pulse of on-resonance repump light immediately before the image. For side-on imaging the imaging beam was detuned by -20~MHz and front-on by -60~MHz to avoid complete absorption and therefore saturation of the measured OD, as well as diffraction effects.

Figure~\ref{fig:cagem_MOTzoo} shows images of the atomic ensemble for various MOT parameters. The images in Figure~\ref{fig:cagem_MOTzoo}(a) were taken of a cloud with all the optimisation protocols described in Sections~\ref{sec:cagem_static} and \ref{sec:cagem_dynamic}, with approximately $4 \cdot 10^9$ atoms present. The peak, front-on OD was measured to be 1000 on resonance with the $F=2\rightarrow F'=3$ transition. This was achieved by detuning the imaging beam approximately $60$~MHz for both positive and negative detunings to ensure no diffraction effects were present and using Equation~\ref{eq:cagem_odscaling}. Two averaged uncalibrated OD cross-sections, taken across ten $y$-plane slices, for these detunings are shown in Figures~\ref{fig:cagem_odcal}(b) and (c). The scaled peak ODs are 900 and 1100 for minus and plus detunings respectively. Averaging the maximum peak height of the individual traces gives over 1000 for both plus and minus detuning. The drop in the average cross-sections can therefore be attributed to slight movement of the MOT between images, with a narrower peak seen for the negative detuning. The temperature of this ensemble, determined by measuring the width of the expanding cloud 5, 10 and 15~ms after the fields were turned off, is approximately 200~$\mu$K in all directions. 

Without the spatial dark spot, Figure~\ref{fig:cagem_MOTzoo}(b)(i) looks very similar to Figure~\ref{fig:cagem_MOTzoo}(a)(i). However, the front-on image in Figure~\ref{fig:cagem_MOTzoo}(b)(ii) shows that the maximum OD drops to 800 with $5 \cdot 10^9$ atoms. Figure~\ref{fig:cagem_MOTzoo}(c) shows the 2D MOT created without turning on the axial coils. This ensemble is much longer in the $z$ direction than the initial cloud (in fact so long that our imaging beam was not large enough to capture it all), but also contains only a quarter of the atoms. This meant that the peak OD dropped to about the same as that without the dark spot. The temperature in the $x$ and $y$ directions for the 2D MOT was approximately 100~$\mu$K but, as there is no trapping in the $z$ direction, the temperature here was much larger. 

Figure~\ref{fig:cagem_MOTzoo}(iv) shows the ensemble without ramping up the 2D MOT magnetic field at the end of the run. This leads to a less-compressed cloud in all three dimensions, though mainly in the $x$-$y$ plane, and therefore a much lower OD of approximately 500. Though there is less magnetic field compression, the temperature of this cloud is the same as the original cloud in the $x$-$y$ plane, with $T = 200$~$\mu$K. However, in the $z$ direction the cloud barely expands over 15~ms.

Finally, Figure~\ref{fig:cagem_MOTzoo}(v) shows the cloud from Figure~\ref{fig:cagem_MOTzoo}(i) expanded for 3~ms as opposed to 500~$\mu$s. The temperature of this cloud will be the same as the initial cloud and, due to the expansion, the peak OD has fallen to 400, though the atom number is still around $4 \cdot 10^9$.\\

\section{Gradient Echo Memory}
\label{sec:cagem_gem}

\subsection{GEM Introduction}
\label{sec:cagem_gemtheory}

\begin{figure}[!ht]
\begin{center}
\includegraphics[width=\columnwidth]
{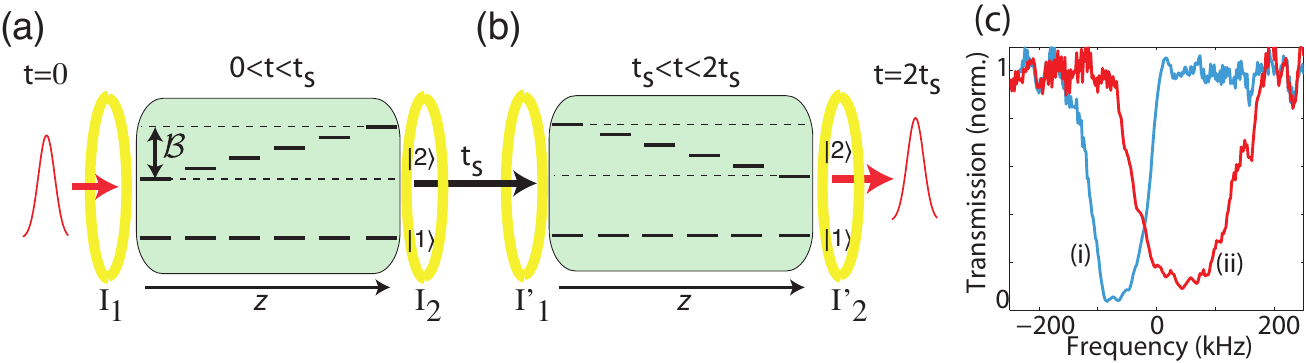}
\caption{Gradient Echo Memory Scheme. (a) Storage: a pulse enters an ensemble of atoms with levels $\left|1\right\rangle$ and $\left| 2 \right\rangle$ at time $0$ where the memory bandwidth $\mathcal{B}=\eta l$. The gradient $\eta$ is created by a pair of coils with currents $I_2 > I_1$. (b) Retrieval: at time $t_s$ the gradient is switched, with $I'_1 = I_2$ and $I'_2 = I_1$ and at time $2t_s$ the echo exits the memory. (c) Averaged and demodulated Raman absorption lines showing broadening due to (i) input gradient and (ii) output gradient.}
\label{fig:cagem_gem}
\end{center}
\end{figure}

As its name suggests, the key to GEM is a linear frequency gradient $\eta$ placed along an ensemble of two-level absorbers, as illustrated in Figure \ref{fig:cagem_gem}. This makes GEM a frequency encoding memory, with information being stored as its spatial Fourier transform along the memory. The details of the scheme are covered in depth in previous papers \cite{Hetet2008z,Hosseini2009,Hosseini2011c,Hosseini2011a,Hosseini2012}.  

The bandwidth of the system is given by $\mathcal{B} = \eta l$, where $l$ is the length of the memory. When using alkali atoms, GEM is implemented using a ground state coherence, in our case between the $F=1$ and $F=2$ hyperfine states. These are coupled using a strong off-resonant coupling beam, as shown in Figure~\ref{fig:cagem_setup}, to make an ensemble of quasi-2-level atoms.  The storage efficiency is determined by the off-resonance broadened Raman line to be
\begin{equation}
\epsilon_s = 1-exp\left(-2 \pi \frac{\mathrm{OD}_{res}}{\mathcal{B'}}\frac{\Omega_c^2}{\Delta^2}\right),
\label{eq:cagem_eff}
\end{equation}
where $\Omega_c$ is the Rabi frequency of the coupling field and $\mathcal{B}'$ is the bandwidth normalised by the excited-state decay rate $\gamma$. To recall the light the gradient is reversed, with $\eta \rightarrow -\eta$. This causes a time reversal of the initial absorption process and the emission of a photon echo from the memory in the forward direction at time $2t_s$, where $t_s$ is the time between pulse arrival and gradient reversal. The monotonicity of the gradient ensures that no light is re-absorbed as it leaves the memory and, as the process is symmetric, the recall efficiency is the same as the storage efficiency, giving a maximum total memory efficiency of $\epsilon_t = \epsilon_s^2$. This does not take into account any decoherence that may go on inside the memory. The detuned Raman absorption nature of GEM (i.e. $\mathrm{OD}_{Ram} \propto (\Omega_c^2/\Delta^2) \mathrm{OD}_{res}/\mathcal{B}$, where $\Omega_c/\Delta \ll 1$) means it requires a much higher OD than a transmissive memory like EIT.  Measurements of the broadened Raman line are shown in Fig.~\ref{fig:cagem_gem}(c) for both the storage and recall gradients.  These measurements show that our writing gradient has a bandwidth of around 50~kHz and the recall gradient has a bandwidth of around 100~kHz. \\

\subsection{GEM Using Cold Atoms}
\label{sec:cagem_cagem}

In our experiment the probe, coupling and local oscillator (LO) fields are all derived from the same laser. The laser can either be locked to the D1 $F=2 \rightarrow F'=2$ transition, using saturated absorption spectroscopy, or placed near this transition, with the frequency being stabilised with a reference cavity. The probe and LO fields are shifted by 6.8 GHz to be resonant with the D1 $F=1 \rightarrow F'=2$ transition using a fibre-coupled electro-optic modulator. All three fields pass through separate acousto-optic modulators to allow for fine frequency adjustment, as well as gating and pulsing. The experimental set-up for the cold atom GEM experiment is shown in Figure \ref{fig:cagem_setup}(b).\\

The reversible frequency gradients for GEM were created using two coils with a radius of 75~mm placed 70~mm apart around the centre of the MOT. These coils were driven with between 2 and 4~A. We also required that the gradient be switched quickly compared to the storage time and with minimal cross-coupling and oscillation after switching. Using a home-built actively stabilised MOSFET-based switch we achieved switching in 1.5~$\mu$s to within 1\% of the desired current \cite{Robins2012b}.\\

\begin{figure}[!ht]
\begin{center}
\includegraphics[width=.5\columnwidth]
{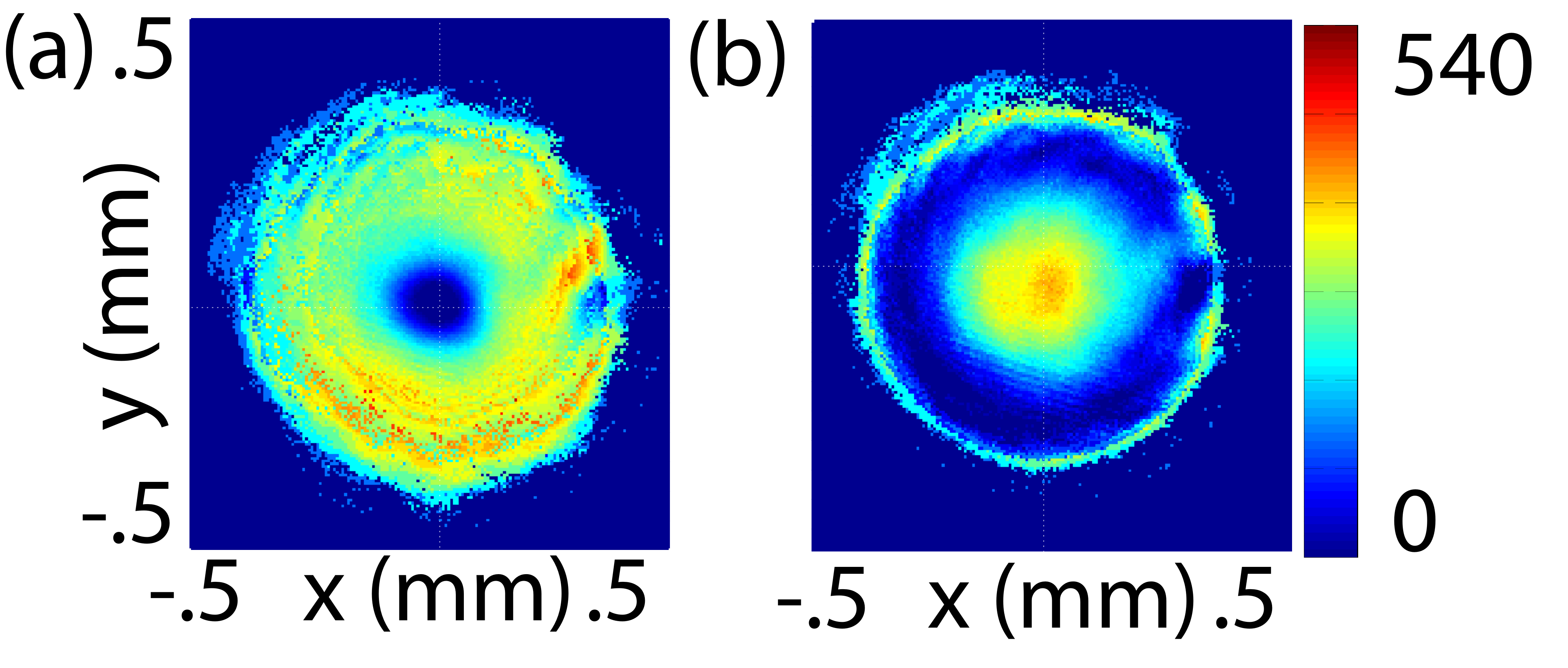}
\caption{Probe Diffraction. Imaging of probe absorption through MOT for (a) -40~MHz and (b) +40~MHz. Colour bar shows OD scale.}
\label{fig:cagem_diff}
\end{center}
\end{figure}

The probe beam was focussed into the MOT to access the highest optical depths available in the atom cloud.  Using lens L2 we formed a waist of 50~$\mu m$ in the atom cloud. Using a CCD we imaged the probe beam transmission as a function of detuning from the $F'=2$ excited state. Two of these images are shown in Figure \ref{fig:cagem_diff}.  We observe that the high density of the atom cloud causes lensing of the probe beam leading to changes in the image as a function of frequency.  These images, combined with additional data from measurements made using an avalanche photodiode and weak probe pulses, indicate a resonant probe optical depth of around 300 on the D1 $F=1 \rightarrow F'=2$ transition. 

To combine the probe and coupling fields we used a non-polarising 50:50 beam-splitter ($BS1$ in Figure \ref{fig:cagem_setup}). Unlike the probe, we want the coupling field to have a large diameter, to cover the entire ensemble with minimal variation in intensity. We use L1 to make a telescope with L2 (the focusing lens for the probe) to have a collimated coupling field with a waist of 1.25~mm. The reason we use a BS rather than a PBS is that it allows us to vary the probe and coupling polarisations, to which GEM is very sensitive \cite{Hosseini2012}. After the MOT we filter the coupling field from the probe using lens L3 and a rectangular pinhole, with dimensions of $100 \times 200$~$\mu$m$^2$. This removes over 99\% of the coupling field and has a probe transmission efficiency of 90\%. The mode selective heterodyne detection then sees no trace of the coupling field.\\

\begin{figure}[!ht]
\begin{center}
\includegraphics[width=\columnwidth]
{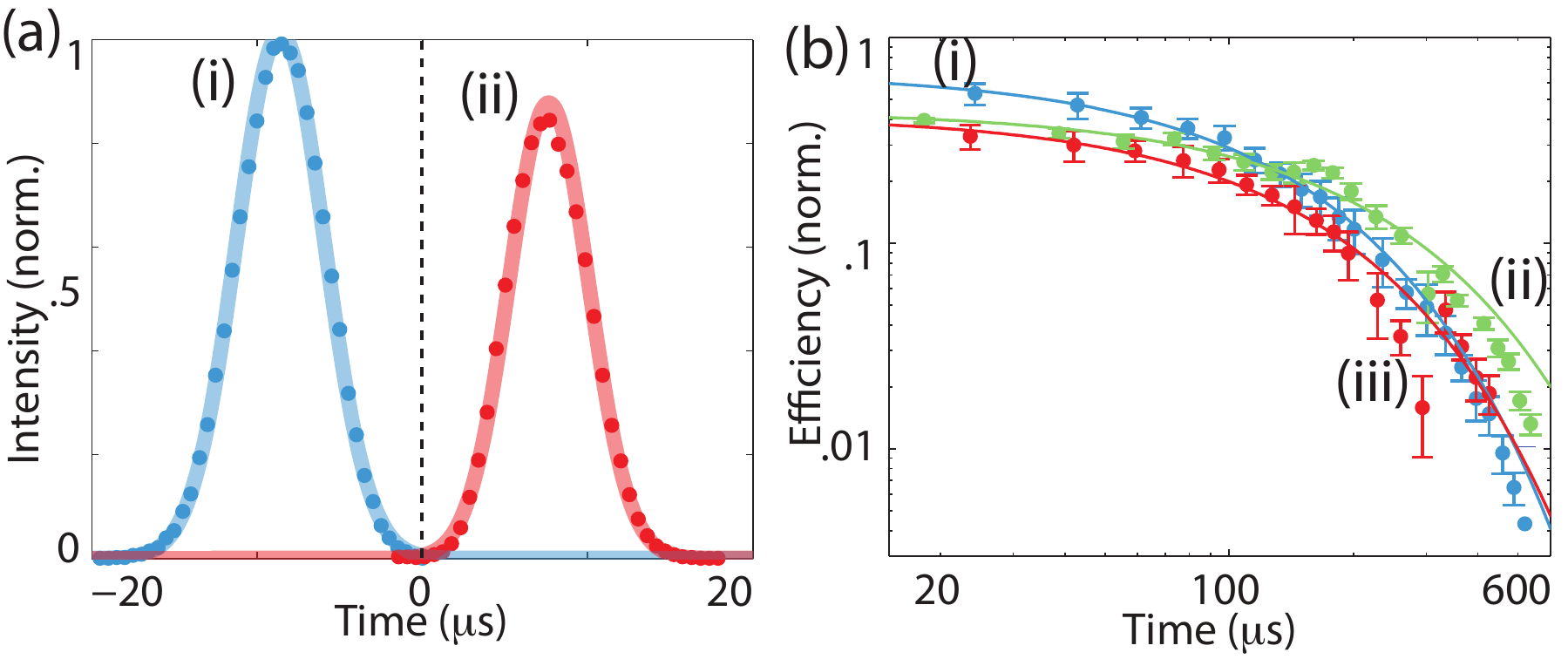}
\caption{GEM with Cold Atoms. (a) High efficiency demodulated and squared heterodyne data: (i) input pulse (blue), and (ii) echo (red) at $80\pm2$\% total efficiency. Points correspond to digitally demodulated data, averaged over 10 (17) traces for input (echo) and squared, error from standard deviation. Lines correspond to Gaussian fit. Dashed line indicates magnetic field switching at $t=0$. (b) Log-log plot of memory efficiency as a function of storage time for (i) MOT shown in Figure~\ref{fig:cagem_MOTzoo}(a) (blue), (ii) MOT shown in Figure \ref{fig:cagem_MOTzoo}(e) (green), and (iii) MOT shown in Figure \ref{fig:cagem_MOTzoo}(c) (red).}
\label{fig:cagem_results}
\end{center}
\end{figure}

For the highest efficiency memory we used the MOT shown in Figure~\ref{fig:cagem_MOTzoo}(a) with a 480~ms load time. After this time the fields are ramped down for 20~ms, as described in Section~\ref{sec:cagem_dynamic}, and then all MOT fields are turned off. At this point the input gradient is turned on with the two GEM coils. 500~$\mu$s later, after allowing eddy currents generated by the MOT coils to switch off, the coupling field is turned on and then the probe is pulsed into the ensemble. Though the MOT cannot fill completely in the 480~ms load time, by continually running the above sequence and having only a millisecond between turning off the MOT and turning it back on again, the atom number in the MOT saturates after only 5 cycles of the experiment.

We found a Gaussian pulse with a full-width-half-maximum of 10~$\mu$s to be optimal for storage. Longer pulses were affected by MOT decay, while shorter pulses required higher bandwidths which reduced the Raman absorption efficiency (see Equation~\ref{eq:cagem_eff}).  For this pulse length, a one-photon detuning of -250~MHz and approximately 350~$\mu$W in the coupling field (corresponding to a Rabi frequency of 2~MHz), we were able to demonstrate storage with $80\pm2$\% total efficiency. This was measured by squaring the modulated pulse, finding the pulse shape, and integrating and averaging over 10 input and 17 output pulses, the error coming from the standard deviation of these traces. This is shown in Figure~\ref{fig:cagem_results}(a), with the data being digitally demodulated in phase, averaged and then squared to produce the intensity plot. As heterodyne detection is mode sensitive, care was taken to optimise the visibility for the input pulse, so that any change in the mode during storage would lead to a reduction in the measured efficiency.

To compare our memory to theoretical expectations, we need to know about the decay of the information stored in the memory and the optical depth.  To investigate the decay we delayed the gradient switching time to store the pulse for longer periods inside the memory, while turning off the coupling field during the storage window. For the MOT used to obtain the 80\% efficiency above we found an exponential decay with a time constant of 117~$\mu$s. The optical depth can be found directly from the storage data where we observed 2\% leakage of the probe field during the write phase. Assuming symmetric read and write operations, the total efficiency of 20~$\mu$s storage can be estimated as $0.98\times0.98\times e^{-20/117}$=81\%.  We know from measurements of the broadened Raman lines (Fig.~\ref{fig:cagem_gem}(c)) that the write and read stages are not perfectly symmetric, however the observed efficiency is still compatible with these measured bandwidths.  We can also crosscheck this result using Eq.~\ref{eq:cagem_eff}.  Taking the resonant optical depth of 150 (as only half the atoms reside in the $m_F$ state used for the memory), the measured bandwidth of the read and write Raman lines and the decay time, we again estimate 81\% total efficiency for our memory.

To investigate the source of the memory decoherence we also measured the decay of the stored light for two other MOT configurations. Firstly, we added a longer waiting period after switching off the MOT: 3~ms (Figure~\ref{fig:cagem_MOTzoo}(v)) instead of 500~$\mu$s. This avoids any residual magnetic fields caused by eddy currents and resulted in an exponential time constant of 195~$\mu$s, though with a lower initial efficiency due to lower initial OD. As the temperature of clouds (i) and (v) are the same, this indicates that temperature is not limiting the coherence time. This is also apparent from the exponential form of the decay: if it were temperature-limited we would expect a Gaussian decay due to the thermal expansion discussed in Section.~\ref{sec:cagem_mot}.

We also investigated the decay using a cloud without the axial coils (Figure~\ref{fig:cagem_MOTzoo}(c)). The initial efficiency was again lower than with the initial MOT, due to the lower OD and size of the cloud, and the time constant was 133~$\mu$s. This indicates the axial diffusion is not the limiting factor for the initial MOT decay, as the initial MOT had a much lower temperature in the $z$ direction. The investigation of decoherence in this system is the subject of ongoing work.  

\section{Discussion}
\label{sec:cagem_discussion}

While 80\% total efficiency is, to the best of our knowledge, the highest efficiency so far reported for cold atomic ensembles, we have numerous paths for further improvement on this result. As with all atomic memory experiments, optical depth is a necessary condition for high efficiency.  In our case, even with the large optical depth we have achieved, it is still the primary limit to our experiment. Both the probe leakage of approximately 8\% and Raman input and output absorption efficiencies indicate that the best possible total efficiency is 85\%, very close to our observed value. One limiting factor on OD is that, when pumped into the $F=1$ state, approximately half the atoms end up in the $m_F=1$ state used for the memory and half in the $m_F=0$. If we implemented an optical pumping scheme similar to the one in \cite{Wang2007} we anticipate an increase in the OD by a factor of two.

The storage time can also be improved by removing sources of decoherence. One potential factor behind the decoherence could be inhomogeneous background magnetic fields, partly due to eddy currents created by the switch off of the MOT coils. Improving this situation may require improvements to switching electronics and physical redesign of the coil configuration. Atomic diffusion will clearly also play a role, so further reductions in temperature will be advantageous.  One more extreme measure would be to transfer atoms into an optical lattice. The increase in coherence time afforded by such a measure must, however, be balanced against the likely decrease in atom number and thus storage efficiency.

Even without further improvement, our memory provides an excellent high-OD platform for numerous other proof-of-principle experiments.  We have proposed previously that the magnetic field gradient could be replaced with an optical (ac Stark) gradient \cite{Sparkes2010b}. Combining this with spatial light modulators and Pockels cells would give switching down to nanoseconds and fine control over the gradient and therefore allow for precision manipulation of the stored information, as discussed in \cite{Buchler2010} and \cite{Sparkes2012}. Another interesting possibility is to investigate cross-phase modulation \cite{Hosseini2011f}, where the high OD, small interaction volume and long storage times are highly advantageous.

\section{Conclusions}
\label{sec:cagem_conclusions}

In this paper we have developed a cold atomic ensemble with $4 \cdot 10^9$ atoms and a peak optical depth of 1000 on the D2 $F=2 \rightarrow F'=3$ transition, specifically with quantum memory applications in mind. We used this ultra-high optical depth system to demonstrate the gradient echo memory scheme on the D1 line with a total efficiency of up to $80\pm2$\% for pulses with a full-width-half-maximum of 10 $\mu$s, still limited primarily by the optical depth. The decoherence of the system was found to be exponential, with a time constant of 117-195~$\mu$s depending on the MOT parameters used, representing an improvement of a factor of 2-4 over the warm GEM experiment.\\

\ack
Thanks to Colin Dedman for his help designing the fast switching GEM coils. This research was conducted by the \textit{Australian Research Council Centre of Excellence for Quantum Computation and Communication Technology} (project number CE110001027). NPR is supported by an Australian Research Council QEII Fellowship. QG is supported by the AFSOR and the Physics Frontier Center at the NIST/UMD Joint Quantum Institute.

\end{document}